\documentclass[12pt]{article}
\begin{document}
\begin{center}
{\bf Particle production related to the tunneling in false vacuum
decay}

\vspace{0.8cm} {
 Michael Maziashvili }

\vspace{0.5cm}\baselineskip=14pt

\vspace{0.5cm} \baselineskip=14pt {\it Department of Theoretical
Physics, Tbilisi State University, 380028 Tbilisi, Georgia}
\end{center}
\vspace{0.5cm}

\begin{abstract}
Motivated by the work of Mersini, the particle production related
to the tunneling in false vacuum decay is carefully investigated
in the thin-wall approximation. It is shown that in this case the
particle production is exponentially suppressed even when the
momentum is comparable to the curvature scale of the bubble. The
number of created particles is ultraviolet finite.
\end{abstract}
\section{General discussion}
False vacuum decay by the tunneling in the WKB approximation is
described by the $O(4)$-symmetric bounce solution,
$\phi_{b}(\sqrt{\vec{x}\,^2+\tau^2}\,)$. In the limit of small
energy-density difference, $\epsilon$, between the true and false
vacuum the bounce looks like a large four-dimensional spherical
bubble of true vacuum separated by a thin wall from a sea of false
vacuum. The evolution of the system after tunneling is described
by the $O(3,1)$ invariant solution that is obtained by the
analytic continuation of the bounce to the Minkowskian time
$\tau\rightarrow it$, see \cite{Col}. Since the field $\phi$
describing the bubble nucleation end expansion is coupled to
itself and to the fluctuation field, the false vacuum decay
through the barrier penetration is accompanied by the particle
production. Using the formalism developed in \cite{Vach-Vil,HSTY}
for the calculation of particle number the particle production by
the thin-walled bubble was considered in \cite{Me}. However, we
fail to corroborate the main conclusion of \cite{Me} that for the
momentum comparable to the curvature scale of the bubble there is
a strong enhancement of the particle production. In this paper we
would like to point out that, in general, the particle production
in the thin-wall approximation is strongly suppressed. The
calculations have been undertaken to exhibit the exponential
suppression of the particle production, even when the momentum is
comparable to the curvature scale of the bubble, rather than for
actual use in calculation of particle number at all momenta.
Throughout this paper the metric signature in Minkowski space-time
is $(-,+,+,+)$ and units $\hbar=c=1$ are used. The calculations
are present in sec.2.
\section{Particle spectrum}
Throughout this paper we restrict ourselves to the consideration
of thin-wall case. This approximation for the bubble formation is
considered in \cite{Col}. We consider a scalar field $\phi$
defined by the Lagrange density\begin{equation}{\cal L}
=-\frac{1}{2}(\partial_{\mu}\phi)^2-U(\phi).\end{equation} The
potential has the form
\begin{equation}U(\phi)=\frac{\lambda}{2}\phi^2(\phi-2a)^2-
\epsilon(\phi^3/2a^3-3\phi^4/16a^4),\label{po}\end{equation} where
$\lambda,~a$ and $\epsilon$ are positive parameters. The potential
(\ref{po}) has a local minimum at $\phi_f=0,~U(0)=0,$ and a global
minimum at $\phi_t=2a,~U(2a)=-\epsilon.$ Bounce solution with a
good accuracy is $\phi_b=a\{1-\tanh(\mu[\rho-R])\}$ where
$\mu=a\sqrt{\lambda}$ and $R=2\mu^3/\epsilon\lambda$. Thickness of
the bubble wall is $L=2/\mu$. For the potential (\ref{po}) the
condition for the validity of the thin-wall approximation is $R/L=
\mu^4/\epsilon\lambda\gg 1$. In other words, in this limiting case
the energy-density difference between the true and false vacuum is
much smaller than the height of the barrier $U(\phi)$. For the
calculation of particle spectrum we use the standard technique
developed in \cite{Vach-Vil,HSTY}. The results relevant for our
discussion can be summarized as follows. The coordinate system
used in the Euclidean region is $(\rho,\chi,\theta,\varphi)$ where
$(\theta,\varphi)$ are usual angle coordinates on two-dimensional
sphere and $(\rho,\chi)$ are related to $r=|\vec{x}|$ and $\tau$
as follows
\begin{eqnarray}r=\rho\sin(\chi),~\tau=-\rho\cos(\chi),\nonumber \\0\leq\chi\leq
\pi/2,~0\leq\rho<\infty.\end{eqnarray} The coordinate system in
the Minkowski region may obtained by replacement $(\rho,\chi)
\rightarrow (-i\rho_{M},-i\chi_{M})$, which yields
\begin{eqnarray}r=\rho_{M}\sinh(\chi_{M}),~t=\rho_{M}\cosh(\chi_{M}),\nonumber
\\0<\chi_{M}<\infty,~0<\rho_{M}<\infty.\end{eqnarray} The basic equation which
governs the fluctuation field $\Phi(t,\vec{x})$ reads
\begin{equation}\left[-\partial_{M}^2-\frac{3}{\rho_{M}}\partial_{M}+\frac{1}{\rho_{M}^2}\hat{L}^2-U''(\phi_b)\right]\Phi=0,\label{1}\end{equation}
where $\partial_{M}$ denotes the partial derivative with respect
to $\rho_{M}$ and $\hat{L}$ is the Laplacian operator on
three-dimensional unit hyperboloid. Since the bounce solution has
$O(3,1)$ symmetry in Minkowski region one can label $\Phi$ by the
eigenvalues of the angular momentum, $(l,m)$ as well as by the
radial momentum, $p$, and write
\begin{equation}\label{an}\Phi=Y_{plm}(\chi,\theta,\varphi)\frac{\psi_{M}(\rho_{M})}{\rho_{M}^{3/2}},\end{equation} where $Y_{plm}$
are eigenfunctions of the operator $-\hat{L}^2$ with eigenvalues
$1+p^2$. After substituting Eq.(\ref{an}) into Eq.(\ref{1}) and
analytically continuing of $\Phi$ to the Euclidean region by
replacement $\rho_{M}\rightarrow i\rho,~\chi_{M}\rightarrow
-i\chi$ one obtains the equation for the fluctuation field in the
under-barrier region
\begin{equation}\left[\partial_{\rho}^2+\frac{(p^2+1/4)}{\rho^2}-U''(\phi_{b})\right]\psi=0,\end{equation}
where $\psi(\rho)\equiv\psi_{M}(i\rho)$. This is a zero-energy
Schr\"{o}dinger
equation\begin{equation}\label{sch}[\partial^2_{\rho}-V(\rho)]\psi=0\end{equation}
of a particle in the potential
\begin{equation}\label{Pot}V(\rho)=-\frac{(p^2+1/4)}{\rho^2}+2\mu^2\{3\tanh^2(\mu[\rho-R])-1\}
+\epsilon\mbox{-term}.\end{equation} Thus, $\psi_{M}(\rho_{M})$ is
given by solving Eq.(\ref{sch}) with $\rho$ replaced by $-i\rho$.
Since in the thin-wall approximation $\phi_{b}(0)$ is very close
to the true vacuum value $2a$ the solution obtained in this way
after the analytical continuation at the turning point $\rho=0$
will have the form
\[\psi_{M}(\rho_{M})=c_1(p)\sqrt{\rho_{M}}\,e^{-\pi p/2}H^{(1)}_{ip}(\omega\rho_{M})+c_2(p)\sqrt{\rho_{M}}\,e^{\pi p/2}H^{(2)}_{ip}(\omega\rho_{M}),\]
where $\omega^2=U''(\phi_{b}(0))$. At $\tau=-\infty$ the field
$\phi$ is in false vacuum state and correspondingly the
fluctuation field $\Phi$ satisfies the vanishing boundary
condition when $\rho\rightarrow\infty$. The spectrum of created
particles $n(p)$ is given by \cite{HSTY}
\begin{equation}n(p)=\frac{1}{(|c_1(p)/c_2(p)|^2-1)}.\end{equation} For detailed description of the general
formalism see \cite{Vach-Vil,HSTY}. \\If $p$ is of order or
smaller than $\mu R$ one can approximate the potential $V$ as
follows
\begin{equation} V=\left\{\begin{array}{ll} 4\mu^2-v/\rho^2 &\mbox{$\rho_{+}<\rho,$}\\
\tilde{v}_3(x+v_2/v_3)^2-k &\mbox{$|x|\leq L,$}\\
4\mu^2-v/\rho^2
&\mbox{$\rho<\rho_{-},$}\end{array}\right.\end{equation} where $
v=p^2+1/4,~x=\rho-R,~
v_1=v/R^2+2\mu^2,~v_2=v/R^3,~v_3=6\mu^4-3v/R^4,~\tilde{v}_3=v_3(6+(p/\mu
R)^2)/24,~k=v_3(v_1/v_3+v_2^2/v_3^2),~\rho_{\pm}=R\pm L$. We have
neglected the contribution coming from the $\epsilon$-term because
at $\phi=0$ this term equals zero and at $\phi\sim a$ this is of
order $\epsilon\lambda/\mu^2\ll\mu^2$. The general solution of
Eq.(\ref{sch}) in regions $|\rho-R|>L$ is written in terms of
modified Bessel functions. For solving this equation we have used
the program package Maple 7. \\In region $\rho<\rho_{-}$ the
solution reads
\begin{equation}\psi=a_1\sqrt{\rho}I_{ip}(2\mu\rho)+a_2\sqrt{\rho}K_{ip}(2\mu\rho).\end{equation}
\\ In region $|x|\leq L$, the exact solution to this equation is
expressed in terms of Whittaker $W,\,M$ functions \cite{Ka}. Using
the program package Maple 7 one finds that in region $|x|\leq L$
the solution has the form
\begin{equation}\psi=a_3W(b,1/4,z^2)/\sqrt{|z|}+a_4M(b,1/4,z^2)/\sqrt{|z|},\end{equation}
where $z=\tilde{v}_3^{1/4}(x+v_2/v_3),\,b=k/4\,\tilde{v}_3^{1/2}$.
For $s=p/\mu R$ by taking into account that in the thin-wall
approximation $(\mu R)^2\ll (\mu R)^4$ one gets
\begin{eqnarray*}k/v_3^{1/2}=(v+2\mu^2R^2+v^2/(6\mu^4R^4-3v))/\sqrt{6\mu^4R^4-3v}\approx\\
((2+s^2)\mu^2R^2+v^2/6\mu^4R^4)/\sqrt{6\mu^4R^4}\approx
(2+s^2)/\sqrt{6},\end{eqnarray*} and
\begin{eqnarray*}(\pm2/\mu+v_2/v_3)v_3^{1/4}=(\pm2/\mu
R+v/(6\mu^4R^4-3v))(6\mu^4R^4-3v)^{1/4}\approx\\(\pm2/\mu
R+v/6\mu^4R^4)\mu R6^{1/4}\approx
\pm2\times6^{1/4}.\end{eqnarray*} It is convenient to introduce
the following notations \begin{eqnarray*}
W_{\pm}(s)=\left.W(b,1/4,z^2)/\sqrt{|z|}\,\right|_{\pm
\tilde{z}},\,W'_{\pm}(s)=\left.\left\{W(b,1/4,z^2)/\sqrt{|z|}\right\}'\right|_{\pm
\tilde{z}},\,\\M_{\pm}(s)=\left.M(b,1/4,z^2)/\sqrt{|z|}\,\right|_{\pm
\tilde{z}},\,M'_{\pm}(s)=\left.\left\{M(b,1/4,z^2)/\sqrt{|z|}\,\right\}'\right|_{\pm
\tilde{z}},\end{eqnarray*} where $\pm \tilde{z}=z(x=\pm
L)\approx\pm 0.1304\times(6+s^2)^{1/4}$ and the prime denotes
differentiation with respect to $z$. These quantities are needed
for the matchings at $\rho_{\pm}$. Using the program package Maple
7 one can estimate $W_{\pm},~W'_{\pm},~M_{\pm},~M'_{\pm}$ at the
desired value of $s$. \\In region $\rho>\rho_{+}$ the solution
satisfying the vanishing boundary condition when
$\rho\rightarrow\infty$ reads
\begin{equation}\psi=a_5\sqrt{\rho}K_{ip}(2\mu\rho).\end{equation}
For our purposes the quantity of immediate interest is
$n(p)=|a_1(p)/a_2(p)\pi|^2$, see \cite{HSTY}. So, one can omit the
common multiplier of $a_1, a_2$ coefficients as well as the common
multiplier of $a_3, a_4$ coefficients. The matching at $\rho_{+}$
gives \begin{equation}\label{match1}\begin{array}{ll}a_3\sim
M'_{+}K_{ip}(2\mu\rho_{+})-M_{+}\{K_{ip}(2\mu\rho_{+})/2\tilde{v}_3^{1/4}\rho_{+}+
K'_{ip}(2\mu\rho_{+})2\mu/\tilde{v}_3^{1/4}\},\\
a_4\sim
W_{+}\{K_{ip}(2\mu\rho_{+})/2\tilde{v}_3^{1/4}\rho_{+}+K'_{ip}(2\mu\rho_{+})2\mu/\tilde{v}_3^{1/4}\}-W'_{+}K_{ip}(2\mu\rho_{+}),\end{array}\end{equation}
the proportionality coefficients are the same for $a_3$ and $a_4$,
here and below the prime denotes differentiation with respect to
the whole argument. The matching at $\rho_{-}$ gives
\begin{eqnarray}\label{match2}a_1\sim \{K_{ip}(2\mu\rho_{-})/2\tilde{v}_3^{1/4}\rho_{-}+K'_{ip}(2\mu\rho_{-})2\mu/\tilde{v}_3^{1/4}\}
\{a_3W_{-}+a_4M_{-}\}-\nonumber\\
K_{ip}(2\mu\rho_{-})\{a_3W'_{-}+a_4M'_{-}\},\nonumber\\
a_2\sim -\{I_{ip}(2\mu\rho_{-})/2\tilde{v}_3^{1/4}\rho_{-}+
I'_{ip}(2\mu\rho_{-})2\mu/\tilde{v}_3^{1/4}\}\{a_3W_{-}+a_4M_{-}\}+\\
I_{ip}(2\mu\rho_{-})\{a_3W'_{-}+a_4M'_{-}\},\nonumber\end{eqnarray}
the proportionality coefficients are the same for $a_1$ and $a_2$.
In the thin-wall approximation $\mu\rho_{\pm}\gg 1$ and for
evaluating the modified Bessel functions of the $2\mu\rho_{\pm}$'s
one can  use the asymptotic formulae. \\For $p\ll\sqrt{\mu R}$ one
can use the following asymptotic expansions , see \cite{AS},
\begin{equation}\label{smallp}\begin{array}{l} I_{ip}(x)\sim \exp(x)\{1+(4p^2+1)/8x+...\}/\sqrt{2\pi x},\\
I'_{ip}(x)\sim \exp(x)\{1-(3-4p^2)/8x+...\}/\sqrt{2\pi
x},\\K_{ip}(x)\sim
\exp(-x)\{1-(4p^2-1)/8x+...\}\sqrt{\pi/2x},\\K'_{ip}(x)\sim
-\exp(-x)\{1+(3-4p^2)/8x+...\}\sqrt{\pi/2x}.
\end{array}\end{equation}
For $p\sim\mu R$ the asymptotic expansions of modified Bessel
functions of purely imaginary order and their derivatives are
derived in \cite{Br}. Motivated in part by the fact that
$K_{ip}(x)$, where $x$ is real variable, is the imaginary part of
$I_{ip}(x)$ up to a multiplicative factor, recent investigations
have been carried out on the real part $I_{ip}(x)$, denoted by
$L_{ip}(x)$,\[L_{ip}(x)=\frac{1}{2}\{I_{ip}(x)+I_{-ip}(x)\}.\] Any
of $L_{ip}(x),K_{ip}(x)$ and $I_{ip}(x)$ can be constructed from
the remaining two functions by the identity
\begin{equation}\label{Id}I_{ip}(x)=L_{ip}(x)-i\frac{\sinh(p\,\pi)}{\pi}K_{ip}(x).\end{equation}
Both $K_{ip}(x)$ and $L_{ip}(x)$ are real valued and even
functions of $p$. \\For $p<x$, introducing $p=x\sin\theta$, where
$0<\theta<\pi/2$, the leading terms of asymptotic expansions are
\cite{Br}
\begin{equation}\label{p<x}\begin{array}{l} L_{ip}(p\csc\theta)\sim \exp(p\,[\cot\theta+\theta])/\sqrt{2p\,\pi\cot\theta}+
\exp(p\,[\pi-\theta-\cot\theta])3p\cot^2\theta/2\pi\,,\\
L'_{ip}(p\csc\theta)\sim
\exp(p\,[\cot\theta+\theta])\sqrt{\cot\theta/2p\,\pi}+\exp(p\,[\pi-\theta-\cot\theta])\tan\theta\sin\theta/3\pi\,,\\
K_{ip}(p\csc\theta)\sim
\exp(-p\,[\cot\theta+\theta])\sqrt{\pi/2p\cot\theta}\,,\\K'_{ip}(p\csc\theta)\sim
-\exp(-p\,[\cot\theta+\theta])\sin\theta\sqrt{\pi\cot\theta/2p}\,.
\end{array}\end{equation}
For $p\approx x$ the leading terms of asymptotic expansions take
the form \cite{Br}
\begin{equation}\label{app}\begin{array}{l} L_{ip}(p)\sim \exp(p\,\pi/2)\{Bi(0)/2^{2/3}p^{1/3}+Bi'(0)/2^{1/3}p^{5/3}70\}\,,\\
L'_{ip}(p)\sim
\exp(p\,\pi/2)\{Bi'(0)/2^{1/3}p^{2/3}-Bi(0)/2^{2/3}p^{4/3}5\}\,,\\
K_{ip}(p)\sim
\pi\exp(-p\,\pi/2)\{2^{1/3}Ai(0)/p^{1/3}+2^{2/3}Ai'(0)/70p^{5/3}\}\,,\\K'_{ip}(p)\sim
\pi\exp(-p\,\pi/2)\{2^{2/3}Ai'(0)/p^{2/3}-2^{1/3}Ai(0)/5p^{4/3}\}\,,
\end{array}\end{equation} where $Ai,~Bi$ are Airy functions.
\\For $p>x$, introducing $p=x\cosh\nu$ and
$\alpha=p(\tanh\nu-\nu)+\pi/4$, where $\nu>0$, the leading terms
of asymptotic expansions are \cite{Br}
\begin{equation}\label{p>x}\begin{array}{l} L_{ip}(p\,\,$sech$\nu)\sim \exp(p\,\pi /2)\sqrt{\coth\nu}\{\cos\alpha+
\sin\alpha\}/\sqrt{2p\,\pi}\,,\\
L'_{ip}(p\,\,$sech$\nu)\sim
\exp(p\,\pi/2)\cosh\nu\{\sin\alpha-\cos\alpha\}/\sqrt{2p\,\pi\coth\nu}\,,\\
K_{ip}(p\,\,$sech$\nu)\sim
\exp(-p\,\pi/2)\sqrt{2\pi\coth\nu}\{\cos\alpha-\sin\alpha\}/\sqrt{p}\,,\\K'_{ip}(p\,\,$sech$\nu)\sim
-\exp(-p\,\pi/2)\sqrt{2\pi}\cosh\nu\{\sin\alpha+\cos\alpha\}/\sqrt{p\coth\nu}\,.
\end{array}\end{equation}
We can now analyze the particle spectrum. Using the leading
asymptotic terms from Eq.(\ref{smallp}) for small values of $p$
one gets
\begin{equation}\label{e1}n(p)\approx16(\mu R)^2\exp(-8\mu R).\end{equation}
Notice that this result is different from that one obtained in
\cite{Me} for small values of $p$. Taking into account that
$W_{\pm},~M_{\pm},~W'_{\pm},~M'_{\pm}$ are real valued functions
from Eqs.(\ref{match1},\ref{match2},\ref{Id}) one obtains
\[a_2\sim Re(a_2)+i\frac{\sinh(p\,\pi)}{\pi}\,a_1,\] and
correspondingly
\begin{equation}\label{su}n(p)=\frac{1}{(\pi Re(a_2)/\,a_1)^2+\sinh^2(p\,\pi)}.\end{equation}
Thus, If $p\gg1$ from Eq.(\ref{su}) one simply concludes that
$n(p)$ is suppressed at least as
\begin{equation}\label{supp}n(p)\sim\exp(-2p\,\pi).\end{equation} As it is clearly seen from Eq.(\ref{su})
$n(p)$ vanishes for that values of $p$ at which $Re(a_2)/a_1$
becomes infinity. Using Eqs.(\ref{p<x},\ref{app},\ref{p>x}) one
can easily estimate the term $Re(a_2)/a_1$ with exponential
accuracy. If $p$ is large, of order $\mu R$, but small than $2\mu
R$ such that $\arcsin(p/x)$ is not very close to $\pi/2$ from
Eq.(\ref{p<x}) one gets
\begin{equation}\label{e2}Re(a_2)/a_1\sim\exp(4p[\cot\theta+\theta]),\end{equation}
where $0<\theta<\pi/2$. For the values of $\theta$ very close to
$\pi/2$ one has to use Eq.(\ref{app}). If $p$ is close to $2\mu R$
using Eq.(\ref{app}) one gets
\begin{equation}\label{e3}Re(a_2)/a_1\sim \exp(4\pi\mu R).\end{equation}
For $p$ of order $\mu R$ such that $p>2\mu R$ from Eq.(\ref{p>x})
one gets
\begin{equation}\label{e4}Re(a_2)/a_1\sim\exp(2p\,\pi).\end{equation}
\\For very large
momentum, assuming for instance $p\gg(\mu R)^2$, one can
approximate $V(\rho)$ as follows
\begin{equation} V=\left\{\begin{array}{ll} 4\mu^2-v/\rho^2 &\mbox{$\rho_{+}\leq\rho,$}\\
-2\mu^2-v/\rho^2 &\mbox{$|x|<L,$}\\ 4\mu^2-v/\rho^2
&\mbox{$\rho\leq \rho_{-}.$}\end{array}\right.\end{equation}
Namely, we have replaced the well of the second term of
Eq.(\ref{Pot}) in region $|x|<L$ by the square well, and if $p$ is
very large one can neglect the contribution coming from such a
deformation of the potential in region $|x|<L$ in comparison with
the term $v/\rho^2$. Now in region $|x|<L$ the general solution of
Eq.(\ref{sch}) is expressed in terms of Bessel functions of purely
imaginary order. With no loss of generality one can choose the
solution in this region to be real valued function. For instance
in this region one can take
\begin{equation}\psi=a_3\sqrt{\rho}Re(J_{ip}(\sqrt{2}\mu\rho))+a_4\sqrt{\rho}Re(Y_{ip}(\sqrt{2}\mu\rho)).\end{equation}
It is clear that for the particle spectrum one gets again the
Eq.(\ref{su}) and simply concludes that for very large momentum
$n(p)$ is suppressed at least as
\begin{equation}n(p)\sim\exp(-2p\,\pi).\end{equation} The asymptotic behavior of
Bessel functions of large purely imaginary order can be found in
\cite{Ba-Er}. Namely, in \cite{Ba-Er} one can find the asymptotic
expansions of $J_{ip}(x)$ and $H^{(1)}_{ip}(x)$,
\begin{equation}\label{Besselp>x}\begin{array}{l}J_{ip}(x)\sim\exp(p\,\pi/2)\exp(i\sqrt{p^2+x^2}-ip\,$arsh$(p/x)-i\pi/4)/\sqrt{2}\pi(p^2-x^2)^{1/4},\\
H^{(1)}_{ip}(x)\sim\sqrt{2}\exp(p\,\pi/2)\exp(i\sqrt{p^2+x^2}-ip\,$arsh$(p/x)-i\pi/4)/\pi(p^2+x^2)^{1/4},\end{array}\end{equation}
and using the standard relation
$Y_{ip}(x)=i\{J_{ip}(x)-H^{(1)}_{ip}(x)\}$ can determine the
asymptotic behavior of $Y_{ip}(x)$. Using
Eqs.(\ref{p>x},\ref{Besselp>x}) one can easily estimate the ratio
$Re(a_2)/a_1$ with exponential accuracy. The matchings at
$\rho_{+},~\rho_{-}$ give $a_{3,4}\sim\exp(-p\,\pi)$ and
$a_1\sim\exp(-p\,\pi/2),~a_2\sim\exp(p\,\pi/2)$ respectively.
Correspondingly, for very large momentum
\begin{equation}\label{e5}Re(a_2)/a_1\sim\exp(2p\,\pi).\end{equation}
Thus, we have explicitly shown that, in general, the particle
production in the thin-wall approximation is exponentially
suppressed. Using Eqs.(\ref{p>x},\ref{Besselp>x}) one can also
evaluate the ultraviolet finiteness  of the number of created
particles. Without carrying out the explicit computation from
these asymptotic formulae one concludes that for very large
momentum the particle production is controlled by the term
$n(p)\sim\exp(-2p\,\pi)\times ($rational function of $p)$. As an
unessential to this problem we do not take into account the terms
enclosed in the braces in Eq.(\ref{p>x}). For this expression the
integral $\int\limits_{p\gg (\mu R)^2}\limits^{\infty}n(p)$ is
certainly convergent. So we infer that the number of created
particles is ultraviolet finite. This completes the review.
\section*{Acknowledgments}
It is our pleasure to acknowledge helpful conversations with
Professors A.\,Khelashvili, G.\,Lavrelashvili and I.\,Lomidze.

\end{document}